\documentclass[twocolumn]{autart}
\usepackage{pifont}    % Enable this line and disable the
\usepackage{amsmath} % assumes amsmath package installed
\usepackage{amssymb}  % assumes amsmath package installed
\usepackage{color}
\usepackage{graphicx}          % Include this line if your
\begin{document}

\begin{frontmatter}

\title{Rapid Lyapunov control of finite-dimensional quantum systems} % Title, preferably not more

%\thanks[footnoteinfo]{The preliminary version of this paper was partly presented at the 2014 Australian Control Conference, 17-18 November 2014, Canberra, Australia. \\
%$^{\dag}$ Corresponding author.}

\author[A1,A2]{Sen Kuang}\ead{skuang@ustc.edu.cn},    % Add the
\author[A2]{Daoyi Dong}\ead{daoyidong@gmail.com},               % e-mail address
\author[A2]{Ian R. Petersen}\ead{i.r.petersen@gmail.com}  % (ead) as shown

\address[A1]{Department of Automation, University of Science and Technology of China, Hefei 230027, PR China}  % Please supply
\address[A2]{School of Engineering and Information Technology, University of New South Wales, Canberra ACT 2600, Australia}             % full addresses here.

                             % keyword list or with the
                                          % help of the Automatica
                                          % keyword wizard

\begin{abstract}                          % Abstract of not more than 200 words.
Rapid state control of quantum systems is significant in reducing the influence of relaxation or decoherence caused by the environment and enhancing the capability in dealing with uncertainties in the model and control process. Bang-bang Lyapunov control can speed up the control process, but cannot guarantee convergence to a target state. This paper proposes two classes of new Lyapunov control methods that can achieve rapidly convergent control for quantum states. One class is switching Lyapunov control where the control law is designed by switching between bang-bang Lyapunov control and standard Lyapunov control. The other class is approximate bang-bang Lyapunov control where we propose two special control functions which are continuously differentiable and yet have a bang-bang type property. Related stability results are given and a construction method for the degrees of freedom in the Lyapunov function is presented to guarantee rapid convergence to a target eigenstate being isolated in the invariant set. Several numerical examples demonstrate that the proposed methods can achieve improved performance for rapid state control of quantum systems.
\end{abstract}

\end{frontmatter}

\section{Introduction}
Quantum control has the potential to play important roles in the development of quantum information technology and quantum chemistry, and has received wide attention from different fields such as quantum information, chemical physics and quantum optics \cite{c1,c2,c3,c4,c41,c51}. Transfer control between quantum states is one of the basic tasks in quantum control. Different control strategies such as optimal control \cite{c5,c6,c43,c44}, adiabatic control \cite{c7,c8}, Lyapunov control methods \cite{c9,c10,c11,c12,c13,c14,c15,c48,c19,c20,c21}, $H_\infty$ and LQG control \cite{c22,c23,c42}, and sliding mode control \cite{c24,c25} have been presented for controller design in quantum systems. Among these control strategies, Lyapunov control methods have been extensively studied for quantum systems due to their simplicity and intuitive nature in the design of control fields \cite{c16,c17,c18,c46,c47}. In Lyapunov control, a Lyapunov function is constructed using information on states or operators related to the quantum system \cite{c9,c10,c11,c12,c13,c14,c15,c48,c19,c20,c21,c16,c17,c18,c46,c47} and the associated control law is designed based on the Lyapunov function (feedback design). Then the control law can be implemented in an open-loop way. From the viewpoint of control theory, one hopes that any system trajectory converges to a desired target state. Unfortunately, the LaSalle invariance principle used in Lyapunov control methods cannot guarantee convergence of any system trajectory to a target state. Some methods such as using implicit Lyapunov functions or switching control methods have been developed to achieve approximate or asymptotic convergence for some specific quantum control tasks (see, e.g., \cite{c10,c12,c15,c19}).

For quantum systems, rapid state control is of importance because a realistic quantum system cannot be perfectly separated from its environment, which will cause a relaxation or decoherence effect. In the context of quantum information processing, rapid control is a basic requirement for performance improvement in quantum computing. In practical applications, robustness has been recognized as a key requirement for the development of quantum technology \cite{c29,c30,c31,c32,c33,c34,c45,c52}.
Rapid control may make the control law more robust to uncertainties in the model or in the control process. Time optimal control methods have been proposed to achieve rapid control for quantum systems \cite{c26,c27,c28}. However, it is very difficult to obtain the optimal control law for general quantum systems \cite{c26,c27,c28}. In \cite{c20}, an optimal Lyapunov design method has been proposed to design a control law for rapid state transfer in quantum systems. Under power-type and strength-type constraints on the control fields, two kinds of Lyapunov control laws were proposed. In particular, the strength-type constraint led to a bang-bang Lyapunov control. In \cite{c21}, the convergence problem for bang-bang Lyapunov control law is further discussed for two-level quantum systems.

The bang-bang Lyapunov control method in \cite{c20,c21} can be used to achieve rapid state control for some practical quantum systems with a high level of fidelity. However, since the control function of bang-bang Lyapunov control is not continuously differentiable, the LaSalle invariance principle cannot be directly used to guarantee convergence. We show that a high-frequency oscillation with an infinitesimal period may occur in bang-bang Lyapunov control, which prevents effective transfer to the target state. Such control fields can also not be realized in the laboratory. In order to achieve rapidly convergent control in state transfer, we propose two classes of new Lyapunov control methods in this paper: switching Lyapunov control and approximate bang-bang Lyapunov control. We first derive a sufficient condition for a two-level system that shows a high-frequency oscillation with an infinitesimal period in bang-bang Lyapunov control. Then we design a switching strategy, i.e., switching between bang-bang Lyapunov control and standard Lyapunov control. For approximate bang-bang Lyapunov control, we design two special control functions that incorporate bang-bang and smoothness properties. These proposed Lyapunov design methods can achieve rapidly convergent control in quantum systems, which is demonstrated by several numerical examples involving a two-level system, a three-level system and a two-qubit superconducting system.

This paper is organized as follows. Section 2 presents the system model, and analyzes the robustness of open-loop quantum control. Section 3 discusses Lyapunov functions with various degrees of freedom, presents several stability results, and develops a construction method for designing the degrees of freedom in the Lyapunov function. A switching strategy between bang-bang and standard Lyapunov control schemes is proposed and the switching condition is investigated in Section 4. In Section 5, we propose two approximate bang-bang Lyapunov control methods. Three numerical examples are presented to demonstrate the performance of the proposed rapid Lyapunov control strategies in Section 6. Conclusions are presented in Section 7. %\textcolor{blue}{In Appendix, we give another switching Lyapunov control strategy, i.e., switching between bang-bang Lyapunov controls with different strengths.}

\textbf{Notation}
\begin{itemize}
\item $i$: the imaginary unit, i.e., $i=\sqrt{-1}$;
\item $[A,B]$: the commutator of the matrices $A$ and $B$, i.e., $[A,B]=AB-BA$;
\item $[A^{(n)},B]$: the repeated commutator with depth $n$, i.e., $[A^{(n)},B]=\underbrace{[A,[A,\cdots,[A}_{n\
\mathrm{times}},B]]]$;
\item $\|A\|$: the induced $2$-norm of the matrix $A$, or the $l_2$-norm of the vector $A$;
\item $A^\mathrm{T}$: the transpose of the matrix $A$;
\item $A^\dag$: the conjugate transpose of the matrix $A$;
\item $\langle \psi|$: the conjugate transpose of the state vector $|\psi\rangle$;
\item $a^*$: the complex conjugate of the complex number $a$;
\item $|a|$: the modulus of the complex number $a$;
\item $\mathbb{R}$: the sets of all real numbers;
\item $\mathbb{C}$: the sets of all complex numbers;
\item $\mathrm{tr}(A)$: the trace of the matrix $A$;
\item $\lambda(A)$: the spectrum of the matrix $A$, i.e., the set of all eigenvalues of $A$;
\item $\Re(a)$: the real part of the complex number $a$;
\item $\Im(a)$: the imaginary part of the complex number $a$.
\end{itemize}

\section{System model and robustness of control}
\subsection{Models of finite-dimensional quantum systems}
Assume that the quantum system under consideration is an $N$-dimensional and controllable closed system \cite{c50}, described by the following Liouville-von Neumann equation:
\begin{equation}\label{1}
\dot\rho(t)=-i\Big[H_0+\sum_{k=1}^mH_ku_k(t),\,\rho(t)\Big],
\end{equation}
where $\rho(t)\in\mathbb{C}^{N\times N}$ is a density matrix describing the state of the system; $H_0$ is the internal Hamiltonian, and $H_k$ is the control Hamiltonian that describes the interaction between the external control fields and the system ($H_0$ and $H_k$ are time-independent Hermitian matrices); and $u_k(t)\;(k=1,\cdots,m)$ are external real-valued control fields. In the energy representation, $H_0$ has a diagonal form, i.e., $H_0=\text{diag}[\lambda_1,\lambda_2,\cdots,\lambda_N]$. We call $\omega_{ab}\triangleq\lambda_a-\lambda_b\;(a,b\in\{1,2,\cdots,N\})$ the transition frequency between the energy levels $\lambda_a$ and $\lambda_b$. Denote $|\lambda_j\rangle$ as the eigenvector of $H_0$ corresponding to the eigenvalue $\lambda_j$, i.e., $|\lambda_j\rangle=[0, \cdots, 0, 1, 0, \cdots, 0]^\mathrm{T}$ where the $j$-th element is 1 and other elements are 0. All $|\lambda_j\rangle, (j=1,\cdots,N)$ form an orthogonal basis of the $N$ dimensional complex Hilbert space $\mathbb{C}^N$.

Since the Hamiltonian $H_0+\sum_{k=1}^mH_ku_k(t)\triangleq H(t)$ is a Hermitian matrix, the evolution of the system (\ref{1}) is unitary. For a given initial state $\rho(0)=\rho_0$, the quantum state at time $t$, $\rho(t)$, can always be written as
\begin{equation}\label{1a}
\rho(t)=U(t)\rho_0U^\dag(t), t\ge0
\end{equation}
where $U(t)$ is a unitary matrix with $U(0)=I$. Equation (\ref{1a}) indicates that the system state $\rho(t)$ at any time always has the same spectrum with the initial state $\rho_0$.

%\textcolor{blue}{According to (\ref{1a}), the evolution state of the system at any time always has the same spectrum with the initial state. This %means that for a given target state $\rho_f$, only those initial states isospectral with $\rho_f$ can be steered to $\rho_f$, which is a necessary %condition that designing the control laws $u_k(t)$ in the system (\ref{1}) to achieve convergence to the target state.}

Using (\ref{1}) and the property of the unitary operator $U(t)U^\dag(t)=U^\dag(t)U(t)=I$, we can undertake a differentiation calculation on both sides of (\ref{1a}) to obtain the Schr\"{o}dinger equation for $U(t)$:
\begin{equation}\label{5a}
\dot U(t)=-i\Big(H_0+\sum_{k=1}^mH_ku_k(t)\Big)U(t)=-iH(t)U(t)
\end{equation}
with $U(0)=I$.

We assume that the control objective is to steer the system to an eigenstate of $H_0$, $\rho_f\triangleq|\lambda_f\rangle\langle\lambda_f|,(f\in\{1,2,\cdots,N\})$. Due to the isospectral evolution property of closed quantum systems and the fact that pure states ($\text{tr}(\rho^2)=1$) and mixed states ($\text{tr}(\rho^2)<1$) have different spectra, we assume that the initial state is a pure state since the target eigenstate $\rho_f$ is a pure state. Moreover, quantum pure states have wide applications in quantum information processing.

%\textcolor{blue}{We make the stipulation: when we discuss convergence to the target eigenstate $\rho_f$, we assume that the initial states are pure states; otherwise, we think that the %initial states are arbitrary. Below, we will also use different terminologies in different places to address this point.}

Also, the following conditions are assumed on the system:
\begin{equation}\label{2}
\omega_{af}\ne\omega_{bf} \ \left(a\ne b;\; a, b\neq f\right);
\end{equation}
\begin{equation}\label{3}
\left(H_{k'}\right)_{jf}\ne0 \ \left(j\ne f;\;\exists\,k'\!\in\!\{1,\,2,\cdots,\,m\}\right).
\end{equation}

Condition (\ref{2}) means that the transition frequencies between the target eigenstate and other eigenstates are distinguishable, and that $H_0$ is non-degenerate, i.e., its all eigenvalues are mutually different. Condition (\ref{3}) implies that there exists a direct coupling between the target eigenstate and any other eigenstate.

\subsection{Robustness of open-loop quantum control}
For a closed quantum system, we may use the model (\ref{1}) to design an open-loop control law and then apply the control law to the practical system. Since any physical system is unavoidably affected by some uncertainties, the robustness of the control law should be considered. Possible perturbations to the quantum system (\ref{1}) include perturbations of the internal Hamiltonian $H_0$, and perturbations in the control Hamiltonian $H_k$, inaccuracy in the control law, and inaccuracy in the initial states. Here, we mainly discuss perturbations in the Hamiltonian. The aim is to show that rapid control can make the control law more robust to uncertainties in the model or in the control process. This is another motivation (besides reducing the relaxation and decoherence effect) to develop rapid Lyapunov control in the following.

We denote the perturbations in the internal and control Hamiltonians as $\delta H_0$ and $\delta H_k$, respectively, where $\delta H_0$ is a real diagonal matrix and $\delta H_k$ are Hermitian matrices. Thus, the internal and control Hamiltonians with perturbations can be written as $\widetilde{H}_0=H_0+\delta H_0$ and $\widetilde{H}_k=H_k+\delta H_k$, respectively. We call the model (\ref{1}) the nominal system and the system with $\delta H_0$ and $\delta H_k$ the perturbed system. Define $\Delta H\triangleq\delta H_0+\sum_{k=1}^m\delta H_ku_k(t)$. Then, the dynamics of the perturbed system can be described as
\begin{equation}\label{4}
\dot{\tilde{\rho}}(t)=-i\Big[\big(H_0+\sum_{k=1}^mH_ku_k(t)\big)+\Delta H,\,\tilde{\rho}(t)\Big],
\end{equation}
where the Hermitian matrix $\Delta H$ is the uncertainty in the Hamiltonian $H(t)$.

We assume that the uncertainty $\Delta H$ satisfies $\|\Delta H\|\leq\varepsilon$ and that the introduction of $\Delta H$ does not break Conditions (\ref{2}) and (\ref{3}), i.e., $\widetilde{\omega}_{af}\ne\widetilde{\omega}_{bf} \,\left(a\ne b\neq f\right)$, where $\widetilde{\omega}_{ab}\triangleq\tilde{\lambda}_a-\tilde{\lambda}_b$ and $\tilde{\lambda}_a$ is the diagonal elements of $\widetilde{H}_0=\text{diag}[\tilde{\lambda}_1,\tilde{\lambda}_2,\cdots,\tilde{\lambda}_N]$; and $(\widetilde{H}_{k'})_{jf}\ne0 \,(j\ne f;\,\exists\,k'\!\in\!\{1,\,2,\cdots,\,m\})$. Now, we examine the effect of the uncertainty $\Delta {H}$ on the quantum system.
\begin{thm}\label{thm1}
We assume $\|\Delta H\|\leq\varepsilon$. For any initial pure state $\rho_0$, the states of the nominal system (\ref{1}) and the perturbed system (\ref{4}) satisfy $\|\,\tilde{\rho}(t)-\rho(t)\,\|\leq\min\{e^{2t\varepsilon }\!-1,\,2\}\;(t\geq0)$. If $\|\,\rho(T)-\rho_f\,\|=\xi_1$ at a finite time $T$, then for an arbitrarily given $\xi\;(\xi_1\leq\xi)$, when $\varepsilon\leq\frac{\ln(1+\xi-\xi_1)}{2T}$, the distance between $\tilde{\rho}(T)$ and the target state satisfies $\|\,\tilde{\rho}(T)-\rho_f\,\|\leq\xi$.
\end{thm}
\begin{pf}
Similar to (\ref{5a}), we write the perturbed system (\ref{4}) in terms of its time evolution operators $\widetilde{U}(t)$ as follows:
\begin{equation}\label{5b}
\dot{\widetilde{U}}(t)=-i\big(H(t)+\Delta H\big)\widetilde{U}(t),\;\widetilde{U}(0)=I.
\end{equation}
Since both $\widetilde{U}(t)$ and $U(t)$ are unitary matrices, we let
\begin{equation}\label{6}
\widetilde{U}(t)=U(t)Q(t).
\end{equation}
Differentiating both sides of (\ref{6}) with respect to $t$ and considering (\ref{5a}) and (\ref{5b}), we have
\begin{equation}\label{7}
i\dot Q(t)=\left(U^\dag(t)\Delta HU(t)\right)Q(t),\;Q(0)=I.
\end{equation}
Define $U^\dag(t)\Delta HU(t)\triangleq\Gamma(t)$. We have $\|\Gamma(t)\|\leq\|\Delta H\|\leq\varepsilon$. The Dyson series solution of (\ref{7}) is the following time-ordered integral
\begin{align}\label{8}
Q(t)&=I+\sum_{n=1}^\infty(-i)^n\int_0^t\text{d}t_1\int_0^{t_1}\text{d}t_2\cdots\nonumber\\
&\int_0^{t_{n-1}}\text{d}t_n\Gamma(t_1)\Gamma(t_2)\cdots\Gamma(t_n)\triangleq I+W(t),
\end{align}
where $t\geq t_1\geq t_2\geq\cdots\geq t_n\geq0$. Considering $\|\Gamma(t)\|\leq\varepsilon$, we have
\begin{eqnarray}\label{9}
\|W(t)\|=\|W^\dag(t)\|\leq\sum_{n=1}^\infty\varepsilon^n\int_0^t\text{d}t_1\int_0^{t_1}\text{d}t_2\cdots\nonumber\\
\int_0^{t_{n-1}}\text{d}t_n=\sum_{n=1}^\infty\varepsilon^n\frac{t^n}{n!}=e^{t\varepsilon}\!-1.
\end{eqnarray}
For any initial state $\rho(0)=\rho_0$, we have
\begin{eqnarray}\label{10}
\|\Delta\rho(t)\|&\triangleq &\|\tilde{\rho}(t)-\rho(t)\|\nonumber\\
&=&\|\,U(t)Q(t)\rho_0Q^\dag(t)U^\dag(t)-U(t)\rho_0U^\dag(t)\|\nonumber \\
&=&\|Q(t)\rho_0Q^\dag(t)-\rho_0\|\nonumber\\
&=&\|\rho_0W^\dag(t)+W(t)\rho_0+W(t)\rho_0W^\dag(t)\|\nonumber\\
&\leq&\|\rho_0W^\dag(t)\|+\|W(t)\rho_0\|+\|W(t)\rho_0W^\dag(t)\|\nonumber\\
&\leq&\|W^\dag(t)\|+\|W(t)\|+\|W(t)\|^2\nonumber\\
&\leq&e^{t\varepsilon}\!-1+e^{t\varepsilon}\!-1+(e^{t\varepsilon}\!-1)^2\nonumber\\
&=& e^{2t\varepsilon}\!-1.
\end{eqnarray}
Considering $\|\tilde{\rho}(t)-\rho(t)\|\leq 2$, we have
\begin{equation}\label{11}
\|\Delta\rho(t)\|\leq\min\{e^{2t\varepsilon}\!-1, 2\}.
\end{equation}
For the perturbed system (\ref{4}), when $\|\rho(T)-\rho_f\|=\xi_1$,
\setlength{\arraycolsep}{1.5pt}
\begin{align}\label{12}
\|\tilde{\rho}(T)-\rho_f\|
&\leq\|\Delta\rho(T)\|+\|\rho(T)-\rho_f\|\nonumber\\
&\leq e^{2T\varepsilon}\!-1+\xi_1.
\end{align}
Hence, when $\varepsilon\leq\frac{\ln(1+\xi-\xi_1)}{2T}$, we have $\|\tilde{\rho}(T)-\rho_f\|\leq\xi$.\quad$\blacksquare$
\end{pf}
%\begin{rem}\label{rem1}
Theorem \ref{thm1} can be regarded as a generalization to quantum systems of the continuous dependence on parameters of solutions to differential equations.
Theorem \ref{thm1} shows that, for given $\xi$ and $\xi_1$, if the nominal system (\ref{1}) can approach the target state within a shorter time period, the perturbed system (\ref{4}) can tolerate larger perturbations when guaranteeing given performance. That is to say, a rapidly convergent control for the nominal system (\ref{1}) may lead to improved robustness. This paper develops rapidly convergent Lyapunov control methods for the system (\ref{1}).

\section{Lyapunov quantum control and stability}
\subsection{Lyapunov control design}
Consider the following Lyapunov function:
\begin{equation}\label{13}
V=\text{tr}\left(P\rho(t)\right),
\end{equation}
where $P$ is a positive semi-definite Hermitian operator that needs to be constructed for completing a given control task.

The time derivative of the Lyapunov function (\ref{13}) is calculated as
\setlength{\arraycolsep}{1.5pt}
\begin{eqnarray}\label{14}
\dot V&=&\text{tr}(P\dot\rho)\nonumber\\
&=&\text{tr}\left(-i\rho\left[P,\;H_0\right]\right)+\sum_{k=1}^mu_k(t)\text{tr}\big(-i\rho\left[P,\;H_k\right]\big).
\end{eqnarray}
We design the control laws by guaranteeing $\dot{V}\leq0$ in (\ref{14}). Considering the fact that $\mathrm{tr}(-i\rho[P,H_0])$ in (\ref{14}) is independent of the control field $u_k(t)$ while $P$ is an unknown Hermitian matrix to be constructed, we let
\begin{equation}\label{15}
[P,\,H_0]=0.
\end{equation}
Since the diagonal matrix $H_0$ is non-degenerate, (\ref{15}) implies that $P$ is also a diagonal matrix. We denote $P\triangleq\text{diag} [p_1,p_2,\cdots,p_N]$.
Using (\ref{15}), (\ref{14}) can be written as
\begin{equation}\label{16}
\dot V=\sum_{k=1}^mu_k(t)T_k(t),
\end{equation}
where $T_k(t)\triangleq\text{tr}(-i\rho(t)[P,H_k])$. For notational simplicity, we also denote $T_k(t)$ as $T_k$ hereinafter.

Thus, by guaranteeing $\dot{V}\leq0$ in (\ref{16}), we design a control law with the following general form:
\begin{equation}\label{17a}
u_k(t)=f_k(T_k), (k=1,2,\cdots,m)
\end{equation}
where the control function $f_k(\cdot)$ satisfies: 1) $f_k(x)\;(x\in\mathbb{R})$ is continuously differentiable with respect to $x$; 2) $f_k(0)=0$; and 3) $f_k(x)\cdot x\leq0$. In particular, we call the following control law the standard Lyapunov control in this paper:
\begin{equation}\label{17b}
u_k(t)=-K_kT_k(t), (k=1,2,\cdots,m)
\end{equation}
where the control gain $K_k>0$ is used to adjust the amplitude of the control field $u_k(t)$.

\subsection{General stability results}
The control law (\ref{17a}) means that the whole ``closed-loop'' system is a nonlinear autonomous system. We use the LaSalle invariance principle to analyze the stability of the system. The LaSalle principle ensures that the system (\ref{1}) with the control fields (\ref{17a}) necessarily converges to the largest invariant set $E$ contained in $M\triangleq\{\rho:\dot V(\rho)=0\}$.

Assume $\bar{\rho}\in E$ and let $\rho(0)=\bar{\rho}$. The invariance property guarantees that $\dot V(\rho(t))=0\;(t\geq0)$, which holds when $u_k(t)=0\;(k=1,\cdots,m)$, i.e.
\begin{equation}\label{21}
T_k=\text{tr}(-i\rho(t)[P,\,H_k])=0,\ (k=1,\cdots,m).
\end{equation}
Substituting the solution of $\dot\rho(t)=-i[H_0,\rho(t)])$ into (\ref{21}) and using (\ref{15}), one has
\begin{align}\label{22}
&\;\text{tr}\left(e^{-iH_0t}\bar{\rho}e^{iH_0t}[P,\,H_k]\right)=\text{tr}\left(e^{iH_0t}H_ke^{-iH_0t}[\bar{\rho},\,P]\right)\nonumber\\
&\;=\text{tr}\Big(\sum_{n=0}^\infty\frac{\big[(iH_0t)^{(n)},\,H_k\big]}{n!}[\bar{\rho},P]\Big)\nonumber\\
&\;=\sum_{n=0}^\infty\frac{-i^nt^n}{n!}\text{tr}\big([H_0^{(n)},H_k][P,\,\bar{\rho}]\big)=0.
\end{align}

Since the time function sequence $1,t,t^2,\cdots$ is linearly independent, and $H_0$ and $P$ are diagonal, we have
\begin{eqnarray}\label{23}
&&\text{tr}\Big(\big[H_0^{(n)},\,H_k\big][P,\,\bar{\rho}]\Big)=\text{tr}\Big(\big(\omega_{jl}^n(H_k)_{jl}\big)\big((p_j-p_l)\bar{\rho}_{jl}\big)\Big)\nonumber\\
&&=0,\ (k=1,\cdots,m;\,n=0,1,2,\cdots).
\end{eqnarray}
Since $H_k$ and $\bar\rho$ are Hermitian matrices, (\ref{23}) reduces to
\setlength{\arraycolsep}{1.5pt}
\begin{align}\label{24}
&\sum_{j<\,l}\Big(\omega_{jl}^n(H_k)_{jl}(p_l-p_j)\bar{\rho}_{lj}+\omega_{lj}^n(H_k)_{jl}^*(p_j-p_l)\bar{\rho}_{lj}^*\Big)\nonumber\\
&=0,(k=1,\cdots,m;\,n=0,1,2,\cdots).
\end{align}
For even and odd $n$, (\ref{24}) has the following forms, respectively:
\setlength{\arraycolsep}{1.5pt}
\begin{align}
\sum_{j<\,l}\Im\Big(\omega_{jl}^n(H_k)_{jl}(p_l-p_j)\bar{\rho}_{lj}\Big)=0,&\ (n=0,2,\cdots);\label{25}\\
\sum_{j<\,l}\Re\Big(\omega_{jl}^n(H_k)_{jl}(p_l-p_j)\bar{\rho}_{lj}\Big)=0,&\ (n=1,3,\cdots).\label{26}
\end{align}
We denote $FN=N(N-1)-2$, and define
\begin{align}\label{27}
\xi_k\triangleq[(H_k)_{12}\bar{\rho}_{21},\cdots,(H_k)_{1N}\bar{\rho}_{N1},
(H_k)_{23}\bar{\rho}_{32},\cdots,\nonumber\\
(H_k)_{2N}\bar{\rho}_{N2},\cdots,(H_k)_{N-1,N}\bar{\rho}_{N,N-1}]^\text{T},
\end{align}
\begin{eqnarray}\label{28}
\mathcal{M}&\triangleq&\left[\begin{array}{cccccccc}
   1&\cdots&1&1&\cdots&1&\cdots&1
\\\omega_{12}^2&\cdots&\omega_{1N}^2&\omega_{23}^2&\cdots&\omega_{2N}^2&\cdots&\omega_{N-1,N}^2
\\\omega_{12}^4&\cdots&\omega_{1N}^4&\omega_{23}^4&\cdots&\omega_{2N}^4&\cdots&\omega_{N-1,N}^4
\\\vdots&\cdots&\vdots&\vdots&\cdots&\vdots&\cdots&\vdots
\\\omega_{12}^{FN}&\cdots&\omega_{1N}^{FN}&\omega_{23}^{FN}&\cdots&\omega_{2N}^{FN}&\cdots&\omega_{N-1,N}^{FN}
\end{array}\right],
\end{eqnarray}
\begin{align}\label{29}
\mathcal{P}\triangleq\text{diag}\,[p_2-p_1,\cdots,p_N-p_1,p_3-p_2,\cdots,\nonumber\\
p_N-p_2,\cdots,p_N-p_{N-1}],
\end{align}
\begin{equation}\label{30}
\Omega\!\triangleq\text{diag}\,[\omega_{12},\cdots\!,\omega_{1N},\omega_{23},\cdots\!,\omega_{2N},\cdots\!,\omega_{N-1,N}].
\end{equation}
Then (\ref{25}) and (\ref{26}) read
\begin{gather}
\mathcal{MP}\Im(\xi_k)=0,\ (k=1,\cdots,m);\label{31}\\
\mathcal{M}\Omega\mathcal{P}\Re(\xi_k)=0,\ (k=1,\cdots,m).\label{32}
\end{gather}
Since the system (\ref{1}) evolves unitarily, the positive limit set of any evolution trajectory has the same spectrum as its initial state. Thus, the invariant set that the system with the control law (\ref{17a}) will converge to can be characterized in the following theorem.
\begin{thm}\label{thm2}
Given an arbitrary initial pure or mixed state $\rho_0$, and under the action of the control fields (\ref{17a}), the system (\ref{1}) converges to the invariant set $E(\rho_0)=\{\bar\rho:\lambda(\bar\rho)=\lambda(\rho_0); \bar\rho=\bar\rho^\dag; \mathcal{MP}\Im(\xi_k)=0,\mathcal{M}\Omega \mathcal{P}\Re(\xi_k)=0,(k=1,\cdots,m)\}$, where $\lambda(A)$ represents the spectrum of ``$A$", and $\mathcal{M},\mathcal{P},\Omega$ and $\xi_k$ are defined by (\ref{27})-(\ref{30}).
\end{thm}
From Theorem \ref{thm2}, the invariant set that the system converges to is dependent on the transition frequencies of the system, the diagonal values of $P$, the initial state $\rho_0$ and the connectivity of $H_k$. When these factors satisfy some particular conditions, it is possible to obtain a simpler form for the invariant set (see \cite{c18}).

%Theorem \ref{thm2} holds for all initial states. For initial mixed states, any control law cannot drive the system (\ref{1}) to a target eigenstate since the evolution of a closed system keeps its entropy while eigenstates have different entropy from mixed states. Hence, we assume that the initial state is a pure state when we investigate convergence in this paper.

\subsection{Construction of Hermitian operator $P$}
In this subsection, we study the construction method of $P$ to achieve convergence to the target eigenstate $\rho_f$. Thus, we only consider the case of initial pure states. For the system (\ref{1}), all possible initial pure states can be divided into two classes: initial states satisfying either $\text{tr}(\rho_0\rho_f)\ne0$ or $\text{tr}(\rho_0\rho_f)=0$. In this subsection, we give a method for constructing $P$ to achieve convergence to the target state for these two classes of initial states, respectively.

When the initial state $\rho_0$ satisfies $\text{tr}(\rho_0\rho_f)\ne0$, we have the following result.
\begin{thm}\label{thm3}
Consider the system (\ref{1}) satisfying Conditions (\ref{2}), (\ref{3}) and with the control fields (\ref{17a}). Assume that the target eigenstate $\rho_f$ and the initial pure state $\rho_0$ satisfy $\mathrm{tr}(\rho_0\rho_f)\ne0$. If the diagonal elements of $P$ satisfy $p_j=p>p_f\geq0,(j=\{1,2,\cdots,N\}/f)$, then $\rho_f$ is isolated in the invariant set $E(\rho_0)$ and the system state starting from $\rho_0$ necessarily converges to $\rho_f$.
\end{thm}
\begin{pf}
Using Conditions (\ref{2}) and (\ref{3}), we can simplify the invariant set $E(\rho_0)$ in Theorem \ref{thm2}. For convenience of expression, we assume that the target eigenstate is the $N$-th eigenstate of $H_0$, i.e., $\rho_f=\rho_N$. If $p_j=p>p_N\geq0,(j=1,\cdots,N-1)$, we have
$\mathcal{MP}=(p_f-p)
\left[\begin{smallmatrix}
   0&\cdots&1&0&\cdots&1&0&\cdots&1
\\0&\cdots&\omega_{1N}^2&0&\cdots&\omega_{2N}^2&0&\cdots&\omega_{N-1,N}^2
\\0&\cdots&\omega_{1N}^4&0&\cdots&\omega_{2N}^4&0&\cdots&\omega_{N-1,N}^4
\\\vdots&\cdots&\vdots&\vdots&\cdots&\vdots&\vdots&\cdots&\vdots
\\0&\cdots&\omega_{1N}^{FN}&0&\cdots&\omega_{2N}^{FN}&0&\cdots&\omega_{N-1,N}^{FN}
\end{smallmatrix}\right]$
and
$\mathcal{M}\Omega\mathcal{P}=(p_f-p)
\left[\begin{smallmatrix}
   0&\cdots&\omega_{1N}&0&\cdots&\omega_{2N}&0&\cdots&\omega_{N-1,N}
\\0&\cdots&\omega_{1N}^3&0&\cdots&\omega_{2N}^3&0&\cdots&\omega_{N-1,N}^3
\\0&\cdots&\omega_{1N}^5&0&\cdots&\omega_{2N}^5&0&\cdots&\omega_{N-1,N}^5
\\\vdots&\cdots&\vdots&\vdots&\cdots&\vdots&\vdots&\cdots&\vdots
\\0&\cdots&\omega_{1N}^{FN+1}&0&\cdots&\omega_{2N}^{FN+1}&0&\cdots&\omega_{N-1,N}^{FN+1}
\end{smallmatrix}\right]$.
Thus, (\ref{31}) and (\ref{32}) are equivalent to\\
$\left[\begin{smallmatrix}
   1&1&\cdots&1
\\\omega_{1N}^2&\omega_{2N}^2&\cdots&\omega_{N-1,N}^2
\\\omega_{1N}^4&\omega_{2N}^4&\cdots&\omega_{N-1,N}^4
\\\vdots&\vdots&\vdots&\vdots
\\\omega_{1N}^{FN}&\omega_{2N}^{FN}&\cdots&\omega_{N-1,N}^{FN}
\end{smallmatrix}\right]
\cdot\Im\left[\begin{smallmatrix}
(H_k)_{1N}\bar{\rho}_{N1}
\\(H_k)_{2N}\bar{\rho}_{N2}
\\\vdots
\\(H_k)_{N-1,N}\bar{\rho}_{N,N-1}\end{smallmatrix}\right]=0$
and
$\left[\begin{smallmatrix}
  \omega_{1N}&\omega_{2N}&\cdots&\omega_{N-1,N}
\\\omega_{1N}^3&\omega_{2N}^3&\cdots&\omega_{N-1,N}^3
\\\omega_{1N}^5&\omega_{2N}^5&\cdots&\omega_{N-1,N}^5
\\\vdots&\vdots&\vdots&\vdots
\\\omega_{1N}^{FN+1}&\omega_{2N}^{FN+1}&\cdots&\omega_{N-1,N}^{FN+1}
\end{smallmatrix}\right]
\cdot\Re\left[\begin{smallmatrix}
(H_k)_{1N}\bar{\rho}_{N1}
\\(H_k)_{2N}\bar{\rho}_{N2}
\\\vdots
\\(H_k)_{N-1,N}\bar{\rho}_{N,N-1}\end{smallmatrix}\right]=0$, respectively. Using Condition (\ref{2}), one can obtain
$[(H_k)_{1N}\bar{\rho}_{N1},(H_k)_{2N}\bar{\rho}_{N2}, \cdots, (H_k)_{N-1,\,N}\bar{\rho}_{N,\,N-1}]^\text{T}=0$.
Using Condition (\ref{3}), we can obtain the relationship
$[\bar{\rho}_{N1},\bar{\rho}_{N2},\cdots,$ $\bar{\rho}_{N,\,N-1}]^\text{T}=0$. Hence, all states in the invariant set $E(\rho_0)$ are of the form $\bar\rho=\left[\begin{smallmatrix}
\bar{A}&0\\
0&\times
\end{smallmatrix}\right]$, where ``$\times$" represents an arbitrary eigenvalue of the initial state $\rho_0$.

Since $\rho_0$ is a pure state, $\rho_0$ has one eigenvalue 1 and $N-1$ eigenvalues 0. Hence, the states in the invariant set $E(\rho_0)$ have the form of $\bar\rho_1=\left[\begin{smallmatrix}
A_1&0\\
0&1
\end{smallmatrix}\right]$ or $\bar\rho_2=\left[\begin{smallmatrix}
A_2&0\\
0&0
\end{smallmatrix}\right]$, where $A_1$ and $A_2$ are Hermitian matrices. For $\bar\rho_1$, all eigenvalues of $A_1$ are 0, which leads to $A_1=0$, i.e., $\bar\rho_1=\rho_f$. For $\bar\rho_2$, $A_2$ has one eigenvalue 1 with multiplicity 1 and one eigenvalue 0 with multiplicity $N-2$. In other words, $E(\rho_0)=\{\rho_f\}\cup\{\bar\rho_2=\left[\begin{smallmatrix}
A_2&0\\
0&0
\end{smallmatrix}\right]:$ $A_2$ has one eigenvalue 1  and $N-2$ eigenvalues 0$\}\triangleq E_1(\rho_0)\cup E_2(\rho_0)$. It is clear that the target eigenstate $\rho_f$ is isolated in $E(\rho_0)$.

For any initial pure state $\rho_0$ which satisfies $\text{tr}(\rho_0\rho_f)\ne0$, one has $\rho_0\notin E(\rho_0)$ or $\rho_0\in E_1(\rho_0)$. When $\rho_0\in E_1(\rho_0)$, the conclusion naturally holds. When $\rho_0\notin E(\rho_0)$, we have $V(\bar\rho_2 )=p>V(\rho_0)>p_f$. Hence, when $\text{tr}(\rho_0\rho_f)\ne0$ and $p>p_f$, the system (\ref{1}) necessarily converges to the target state $\rho_f$.\quad$\blacksquare$
\end{pf}
When the initial state $\rho_0$ satisfies $\text{tr}(\rho_0\rho_f)=0$, we have $\rho_0\in E_2(\rho_0)$. That is to say, under the construction relation of $P$ in Theorem \ref{thm3}, the control law (\ref{17a}) cannot enable any state transfer. In this case, there exists a $j\in\{1,2,\cdots,N\}/f$ such that $\langle\lambda_j|\rho_0|\lambda_j\rangle\ne0$. Thus, we may use the following switching control to achieve convergence to the target state:
\begin{eqnarray} \label{33}
u_k(t)=\left\{ \begin{aligned}
         &f_k(\sin(\omega_{jf}t)), t\in[0,\,t_0] \\
         &\quad f_k(T_k), \quad\quad t>t_0
         \end{aligned} \ (k=1,\cdots,m),\right.
\end{eqnarray}
where $\omega_{jf}\triangleq\lambda_j-\lambda_f$, and $t_0$ is a small time duration.

When $t_0$ is small, the state $\rho(t_0)$ is not in the invariant set $E(\rho_0)$. If we take $\rho(t_0)$ as a new initial state, then Theorem \ref{thm3} guarantees that the control law (\ref{17a}) can achieve convergence to the target state. Thus, we have the following conclusion.
\begin{thm}\label{thm4}
Consider the system (\ref{1}) satisfying Conditions (\ref{2}), (\ref{3}) and with the switching control (\ref{33}). Assume that the target eigenstate $\rho_f$ and the initial pure state $\rho_0$ satisfy $\mathrm{tr}(\rho_0\rho_f)=0$. If the switching time $t_0$ satisfies $\rho(t_0)\notin E(\rho_0)$ and the diagonal elements of $P$ satisfy $p_j=p>p_f\geq0\;(j=\{1,2,\cdots,N\}/f)$, then the system state starting from $\rho_0$ necessarily converges to $\rho_f$.
\end{thm}
For general continuously differentiable control function (\ref{17a}), the construction relation in Theorems \ref{thm3} and \ref{thm4} ensures convergence to the target eigenstate. Based on the construction relation of $P$, we propose two new methods including switching Lyapunov control and approximate bang-bang Lyapunov control to achieve rapidly convergent Lyapunov control.

\section{Switching between Lyapunov control schemes}
To speed up the control process, Ref. \cite{c20} proposed two design methods for quantum systems with power-type constraints and strength-type constraints such that $\dot{V}$ in (\ref{16}) takes the minimum value at each moment. For the case with strength-type constraints, the ``optimal'' control law is the following bang-bang Lyapunov control:
\begin{equation} \label{18}
u_k(t)=\left\{ \begin{aligned}
         -S,\ (T_k>0) \\
         S\,,\ (T_k<0) \\
         0\,,\ (T_k=0)
         \end{aligned} \quad (k=1, \cdots, m),\right.
\end{equation}
where $S$ is the maximum admissible strength of each control field, i.e., $|u_k(t)|\leq S$.

The bang-bang Lyapunov control in (\ref{18}) makes $\dot{V}$ in (\ref{16}) satisfy $\dot{V}\leq0$, and can speed up completing some quantum control tasks. Especially, the state may move rapidly towards the target state at the early stages of the control \cite{c20}. However, convergence cannot be guaranteed since the control function is not continuously differentiable. Here, we first show that the bang-bang Lyapunov control may lead to a high-frequency oscillation phenomena \cite{c20,c21}, which prevents effective state transfer towards the target state. Then, we propose two classes of switching Lyapunov control strategies to achieve rapidly convergent control, i.e., switching between the bang-bang Lyapunov control and the standard Lyapunov control, and switching between bang-bang Lyapunov control schemes with variable control strengths.

\subsection{High-frequency oscillation in bang-bang Lyapunov control}
In this subsection, we present a sufficient condition for two-level quantum systems that high-frequency oscillation phenomena occur in the bang-bang Lyapunov control \cite{c20}, which can be used to determine switching conditions for the design of switching Lyapunov control. We first give the following definition.

\begin{defn}\label{def1}
The control law (\ref{18}) is said to have a high-frequency oscillation with an infinitesimal period at time $t_{0}$ if $\,\exists\,\epsilon >0$,
$$\inf \{\tau>0: u_{k}(t+\tau)\neq u_{k}(t) \}=0 $$
for all $t\in [t_{0}, t_{0}+\epsilon]$.
\end{defn}
Since the control in (\ref{18}) can take on only one of three constant values (0 and $\pm S$) at any time, the high-frequency oscillation in Definition \ref{def1} means that the control always jumps between these values after an arbitrarily small time duration in the interval $[t_0,t_0+\epsilon]$. Such a control field cannot be realized in practice.

Now consider the system model (\ref{1}) in the case of two energy levels, and denote its internal Hamiltonian as $H_0$ and its control Hamiltonian as $H_1$:
\begin{equation}\label{36}
H_0=\begin{bmatrix}
\lambda_1&0\\
0&\lambda_2
\end{bmatrix},\
H_1=\begin{bmatrix}
0&r\\
r^*&0
\end{bmatrix},
\end{equation}
where $r\ne0$.

We define the first eigenstate $|\lambda_1\rangle\triangleq[1, 0]^\text{T}$ as the excited state, $|\lambda_2\rangle\triangleq[0, 1]^\text{T}$ as the ground state, and
\begin{equation}\label{36c}
\omega_{12}\triangleq\lambda_1-\lambda_2>0.
\end{equation}
Let the excited state $|\lambda_1\rangle$ be the target state. According to Theorem \ref{thm3} or Theorem \ref{thm4}, $P$ can be chosen as $P=\mathrm{diag}\,[p_1,p],(p>p_1)$. Thus, $[P,\,H_1]=(p-p_1)\left[\begin{smallmatrix}
0&-r\\
r^*&0
\end{smallmatrix}\right]$.
From (\ref{18}), $T_1=0$ holds at any zero point of the bang-bang Lyapunov control. For convenience of analysis, in this paper we denote such moments as $\tilde{0}$ to differentiate them from the initial moment 0.
Thus, we have
\begin{align}\label{37}
T_1(\tilde{0})&=-i\text{tr}\big([P,H_1]\rho(\tilde{0})\big)\nonumber\\
&=2(p-p_1)\cdot\Im\big(r^*\rho_{12}(\tilde{0})\big)\nonumber\\
&=0.
\end{align}
Equation (\ref{37}) equals that $r^*\rho_{12}(\tilde{0})\in\mathbb{R}$.
For the two-level system, we have the following result.
\begin{thm}\label{thm5}
Consider the two-level system
\begin{equation}\label{2level}
\dot\rho(t)=-i [H_0+H_1 u_1(t),\;\rho(t) ],
\end{equation}
with the Hamiltonians (\ref{36}) and the bang-bang Lyapunov control (\ref{18}) (where $k=1$). Assume that the initial state of the system is an arbitrary pure state. We denote the state at any zero point  of the control field $\,\tilde{0}$ (i.e., $u_{1}(t=\tilde{0})=0$) as $\rho(\tilde{0})=\left[\begin{smallmatrix}
\rho_{11}(\tilde{0})&\rho_{12}(\tilde{0})\\
\rho_{12}^*(\tilde{0})&\rho_{22}(\tilde{0})
\end{smallmatrix}\right]$. Then, a sufficient condition for the bang-bang Lyapunov  control (\ref{18}) to have a high-frequency oscillation with an infinitesimal period is
\begin{equation}\label{40}
\frac{|r|\left(\rho_{11}(\tilde{0})-\rho_{22}(\tilde{0})\right)}{|\rho_{12}(\tilde{0})|}\geq\frac{\omega_{12}}{S}, \left(\rho_{12}(\tilde{0})\ne0\right).
\end{equation}
\end{thm}
\begin{pf}
Assume that from the state $\rho(\tilde{0})$, a constant control $u_1(t)=u$ acts on the system and lasts to time $t$. Write the state at time $t$ as
$\rho(t)=e^{-i(H_0+H_1u)t}\rho(\tilde{0})e^{i(H_0+H_1u)t}$. We have
\begin{eqnarray}\label{42}
T_1(t)&=&-i\,\text{tr}\big([P,H_1]\rho(t)\big)\nonumber\\
&=&-i\text{tr}\big(e^{i(H_0+H_1u)t}[P,H_1]e^{-i(H_0+H_1u)t}\rho(\tilde{0})\big).
\end{eqnarray}
Denote $\omega_u=\sqrt{\omega_{12}^2+4|r|^2u^2}$, $R_1=\left[\begin{smallmatrix}
0&-r\\
r^*&0
\end{smallmatrix}\right]$, and $R_2=\left[\begin{smallmatrix}
2|r|^2u&-r\omega_{12}\\
-r^*\omega_{12}&-2|r|^2u
\end{smallmatrix}\right]$, then $e^{i(H_0+H_1u)t}[P,H_1]e^{-i(H_0+H_1u)t}$ in (\ref{42}) can be calculated as
\begin{eqnarray}\label{43}
&&e^{i(H_0+H_1u)t}[P,H_1]e^{-i(H_0+H_1u)t}\nonumber\\
&=&[P,H_1]+it[H_0+H_1u,[P,H_1]]\nonumber\\
&&+\frac{(it)^2}{2!}[H_0+H_1u,[H_0+H_1u,[P,H_1]]]\nonumber\\
&&+\frac{(it)^3}{3!}[H_0+H_1u,[H_0+H_1u,[H_0+H_1u,[P,H_1]]]]\nonumber\\
&&+\cdots\nonumber\\
&=&(p-p_1)R_1+it(p-p_1)R_2+\frac{(it)^2}{2!}(p-p_1)\omega_u^2R_1\nonumber\\
&&+\frac{(it)^3}{3!}(p-p_1)\omega_u^2R_2+\frac{(it)^4}{4!}(p-p_1)\omega_u^4R_1\nonumber\\
&&+\frac{(it)^5}{5!}(p-p_1)\omega_u^4R_2+\cdots\nonumber\\
&=&(p-p_1)\cos\left(\omega_ut\right)R_1+\frac{i(p-p_1)}{\omega_u}\sin\left(\omega_ut\right)R_2,
\end{eqnarray}
where we have used the series formula $\sum_{n=0}^\infty\frac{x^{2n+1}}{(2n+1)!}=\frac{e^x-e^{-x}}{2}$ and $\sum_{n=0}^\infty\frac{x^{2n}}{(2n)!}=\frac{e^x+e^{-x}}{2}, (x\in\mathbb{R})$.

Substituting (\ref{43}) into (\ref{42}) gives
\begin{align}\label{44}
T_1(t)&=-i(p-p_1)\cdot\cos(\omega_ut)\cdot\text{tr}(R_1\rho(\tilde{0}))\nonumber\\
&\quad\quad\quad+\frac{(p-p_1)}{\omega_u}\cdot\sin(\omega_ut)
\cdot\text{tr}(R_2\rho(\tilde{0}))\nonumber\\
&=\frac{2(p-p_1)}{\omega_u}\cdot\sin(\omega_ut)\cdot\nonumber\\
&\quad\big[u|r|^2\left(\rho_{11}(\tilde{0})-\rho_{22}(\tilde{0})\right)-\omega_{12}\cdot r^*\rho_{12}(\tilde{0})\big].
\end{align}
Using (\ref{44}), we prove the conclusion in the theorem by contradiction.

For $\rho_{12}(\tilde{0})\ne0$, we first assume that the control field after $\tilde{0}$ is $u=S$ which can last for a given small time duration $\tilde{t}_1$ ($\tilde{t}_1>0$). Then, (\ref{18}) guarantees that $T_1(t)<0$ holds for $t\in(\tilde{0},\tilde{t}_1)$. It follows from the condition (\ref{40}) that $S|r|^2(\rho_{11}(\tilde{0})-\rho_{22}(\tilde{0}))\geq\omega_{12}|r^*\rho_{12}(\tilde{0})|\geq\omega_{12}r^*\rho_{12}(\tilde{0})$. Considering (\ref{44}), we have $T_1(t)\geq0$ holds for $t\in(\tilde{0},\tilde{t}_1)$. Such a contradiction implies that the constant control $u=S$ from $\tilde{0}$ cannot last for any finite time duration $\tilde{t}_1$. Similarly, if the control field after $\tilde{0}$ is $u=-S$ which can last for a small non-zero time duration $\tilde{t}_1$, then (\ref{18}) guarantees that $T_1(t)>0$ holds for $t\in(\tilde{0},\tilde{t}_1)$. It follows from the condition (\ref{40}) that $S|r|^2(\rho_{11}(\tilde{0})-\rho_{22}(\tilde{0}))\geq\omega_{12}|r^*\rho_{12}(\tilde{0})|\geq-\omega_{12}r^*\rho_{12}(\tilde{0})$, i.e., $-S|r|^2(\rho_{11}(\tilde{0})-\rho_{22}(\tilde{0}))-\omega_{12}r^*\rho_{12}(\tilde{0})\leq0$. Considering (\ref{44}), we know that $T_1(t)\leq0$ holds for $t\in(\tilde{0},\tilde{t}_1)$. Such a contradiction implies that the constant control $u=-S$ cannot last for a finite time duration from $\tilde{0}$. For the case of $u=0$, it is straightforward to obtain a contradiction from (\ref{44}).
%Thus, we can conclude that when the condition (\ref{40}) is satisfied, the non-zero constant control must have the second zero point $\tilde{0}_2$ at a certain infinitesimal moment starting %from $\tilde{0}$.

Since $T_{1}(t)$ is a continuous function of $t$, we consider another zero point $\tilde{0}_2$ following $\tilde{0}$. Considering that the Lyapunov function $V=p_1\rho_{11}+p\rho_{22}=p-(p-p_1)\rho_{11},\,(p>p_1)$ satisfies $\dot{V}\leq0$ in the interval $[\tilde{0},\,\tilde{0}_2]$, we have $\rho_{11}(\tilde{0}_2)\geq\rho_{11}(\tilde{0})$. The condition (\ref{40}) implies that $\rho_{11}(\tilde{0})>\rho_{22}(\tilde{0})$, i.e., $\rho_{11}(\tilde{0})>\frac{1}{2}$. Thus, we have $\rho_{11}(\tilde{0}_2)\geq\rho_{11}(\tilde{0})>\frac{1}{2}$. Therefore, $\frac{|r|(\rho_{11}(\tilde{0}_2)-\rho_{22}(\tilde{0}_2))}{|\rho_{12}(\tilde{0}_2)|}=\frac{|r|(2\rho_{11}(\tilde{0}_2)-1)}{\sqrt{\rho_{11}(\tilde{0}_2)(1-\rho_{11}(\tilde{0}_2))}}\geq\frac{|r|(2\rho_{11}(\tilde{0})-1)}{\sqrt{\rho_{11}(\tilde{0})(1-\rho_{11}(\tilde{0}))}}=\frac{|r|(\rho_{11}(\tilde{0})-\rho_{22}(\tilde{0}))}{|\rho_{12}(\tilde{0})|}\geq\frac{\omega_{12}}{S}$ holds. That is to say, the condition (\ref{40}) still holds at the zero point $\tilde{0}_2$. We can conclude that the bang-bang Lyapunov control has a high-frequency oscillation with an infinitesimal period when the condition (\ref{40}) is satisfied.
\quad$\blacksquare$
\end{pf}

According to Theorem \ref{thm5}, when $\rho(\tilde{0})$ satisfies (\ref{40}), the bang-bang Lyapunov control has a high-frequency oscillation with an infinitesimal period. Such a control field with a high-frequency oscillation cannot guarantee effective state transfer for the two-level system as well as it is not physically realizable. When $\rho_{12}(\tilde{0})\ne0$ and
\begin{gather}\label{abc}
\frac{|r|\left(\rho_{11}(\tilde{0})-\rho_{22}(\tilde{0})\right)}{|\rho_{12}(\tilde{0})|}<\frac{\omega_{12}}{S},
\end{gather}
we can show that there exists an appropriate bang-bang Lyapunov control to achieve effective state transfer. For example, when $r^*\rho_{12}(\tilde{0})>0$ and $u=S$, $T_{1}(t)<0$ in (\ref{44}) can last at least for $\frac{\pi}{\omega_{u}}$. Hence, the condition in (\ref{40}) will be used to design the switching control law in the following subsection.

\begin{rem}\label{rem2a}
Since $\omega_{12}>0$, the right-hand side of (\ref{40}) is positive. Hence, $\rho_{11}(\tilde{0})>\rho_{22}(\tilde{0})$ in the left side always holds. This clearly shows that any high-frequency oscillation with an infinitesimal period only may occur when $\rho_{11}(\tilde{0})>\frac{1}{2}$.
For any zero point $\tilde{0}$ of the bang-bang Lyapunov control (\ref{18}), when the state $\rho(\tilde{0})$ satisfies $\rho_{12}(\tilde{0})=0$, a direct calculation of $\,T_{1}$ with only an internal Hamiltonian $H_{0}$ shows that the control field will always be zero and the state transfer will stop. That is to say, the system state in this case is within the invariant set. When $\rho(\tilde{0})$ satisfies (\ref{40}), the control law has a high-frequency oscillation with an infinitesimal period and cannot guarantee convergence toward the target state.
\end{rem}
\begin{rem}\label{rem3}
Theorem \ref{thm5} only considers two-level systems. For general $N$-level systems, the high-frequency oscillation phenomena in bang-bang Lyapunov control may also occur. However, it is very difficult to establish a strict sufficient condition for the high-frequency oscillation phenomena in this case.
\end{rem}

\subsection{Switching design between bang-bang and standard Lyapunov control schemes}
We consider two-level systems. In order to avoid possible high-frequency oscillations with infinitesimal periods in the bang-bang Lyapunov control (\ref{18}), we design the switching control as follows. If the high-frequency oscillation condition in Theorem \ref{thm5} is not satisfied, we apply the bang-bang Lyapunov control law (\ref{18}) to the system; once the condition is satisfied at a certain zero point $\tilde{0}$, we switch to the standard Lyapunov control.

The standard Lyapunov control (\ref{17b}) needs to satisfy the strength constraint, i.e., $|u_1(t)|\leq S$. Calculating $T_1(t)$ gives
\begin{eqnarray}
T_1(t)&=&-i(p-p_1)\mathrm{tr}\!\!\left(
\begin{bmatrix}
\rho_{11}(t)&\rho_{12}(t)\\
\rho_{12}^*(t)&\rho_{22}(t)
\end{bmatrix}
\begin{bmatrix}
0&r\\
-r^*&0
\end{bmatrix}
\right)\nonumber\\
&=&-2(p-p_1)\cdot\Im(r^*\rho_{12}(t)).\label{a35a}
\end{eqnarray}
%Since the evolution state starting from any initial pure state always keeps pure, we write $\rho_{12}(t)$ in (\ref{a35a}) as $\rho_{12}(t)=c_1(t)c_2^*(t)$, where $c_j(t),\,(j=1,\,2)$ is the %$j$-th component of the evolution state column vector. Thus,
We can obtain an estimate of the amplitude of $T_1(t)$ as
\begin{align}\label{35b}
|T_1(t)|&=2(p-p_1)\cdot\Im(r^*\rho_{12}(t))\nonumber\\
&\leq2(p-p_1)\cdot|r\rho_{12}(t)|\nonumber\\
&=2(p-p_1)|r|\sqrt{(1-\rho_{11}(t))\rho_{11}(t)}\nonumber\\
&\leq(p-p_1)|r|.
\end{align}
For (\ref{35b}), when $\rho_{11}(t)\!=\!\frac{1}{2}$, $2(p-p_1)|r|\sqrt{(1\!-\!\rho_{11}(t))\rho_{11}(t)}$ reaches its maximum value. $|T_1(t)|$ may reach its maximum when the initial $\rho_0$ satisfies $\text{tr}(\rho_0\rho_f)<\frac{1}{2}$. To guarantee that the standard Lyapunov control does not exceed the maximum admissible strength for all possible initial states, we choose
\begin{equation}\label{68}
K_1=\frac{S}{|T_1|_{\mathrm{max}}}=\frac{S}{(p-p_1)|r|}.
\end{equation}
Thus, we can design a switching control law as follows:
\begin{equation} \label{69}
u_1(t)=\left\{ \begin{aligned}
         &-S\cdot \mathrm{sgn}(T_1(t)) , (\text{until}\ (\ref{40})\ \text{holds}) \\
         &-K_1\cdot T_1(t), \;\mathrm{otherwise}
         \end{aligned} \right.
\end{equation}
where $K_1=\frac{S}{(p-p_1)|r|}$. Note that switching between the bang-bang Lyapunov control and the standard Lyapunov control may only occur at zero points of $T_1(t)$. From the initial time $t=0$, the bang-bang Lyapunov control should be first used unless $T_{1}(0)=0$.

\begin{rem}
Observing the high-frequency oscillation condition for two-level systems in Theorem \ref{thm5}, it is clear that reducing the bang-bang Lyapunov control strength can avoid high-frequency oscillations. This observation tells us that we can also develop a new switching design strategy involving switching between bang-bang Lyapunov controls with different control bounds. Such a switching strategy is outlined in the Appendix.
\end{rem}

\subsection{Stability of switching Lyapunov control}
Based on the construction relation of $P$ in subsection 3.3, bang-bang Lyapunov control can speed up the control process, but cannot guarantee convergence to the target state; while the standard Lyapunov control can guarantee convergence. Therefore, dependent on different initial states, the switching design strategy in subsection 4.2 can achieve rapidly convergent control.
\begin{thm}\label{thm6}
Consider the two-level system (\ref{2level}) with the Hamiltonians (\ref{36}). Assume that the target state $\rho_f$ is the excited state $\rho_1=|\lambda_1\rangle\langle\lambda_1|$, and that the initial state $\rho_0$ is an arbitrary pure state. The diagonal elements of $P$ satisfy $p_2=p>p_f=p_1\ge0$. Then, the following conclusions hold:\\
$\mathrm{i})$ the largest invariant set of the system with the switching Lyapunov control (\ref{69}) is $E'=\{\rho_f\}\cup\{\rho_2\}$, where $\rho_2\triangleq|\lambda_2\rangle\langle\lambda_2|$;\\
$\mathrm{ii})$ when $\rho_0$ satisfies $\mathrm{tr}(\rho_0\rho_f)\ne0$, with the switching Lyapunov control (\ref{69}), the system trajectory starting from $\rho_0$ necessarily converges to $\rho_f$;\\
$\mathrm{iii})$ when $\rho_0$ satisfies $\mathrm{tr}(\rho_0\rho_f)=0$, the initial control $u_1 (t)=S \sin⁡(\omega_{21} t)\,(t\in[0,t'])$ is first used, where $\omega_{21}$ $\triangleq\lambda_2-\lambda_1$ and $t'$ is a small positive number satisfying $\mathrm{tr}(\rho(t')\rho_f)\ne0$; then, with the switching Lyapunov control (\ref{69}), the system trajectory starting from $\rho(t')$ necessarily converges to $\rho_f$.
\end{thm}
\begin{pf}
Conclusion $\mathrm{i})$. According to the proof of Theorem \ref{thm3} ($k=1$), for the initial state $\rho_0$ and the target state $\rho_f$, the largest invariant set of the two-level system (\ref{2level}) only under the action of the standard Lyapunov control is $E'=\{\rho_f\}\cup\{\rho_2=\left[\begin{smallmatrix}
0&0\\
0&1
\end{smallmatrix}\right]\}$.
%(Note that the target state in the proof of Theorem \ref{thm3} is assumed to be the $N$-th eigenstate, while the target state here is the first eigenstate).
It can be verified that when the system is in the state $\rho_f$ or $\rho_2$, the switching Lyapunov control $u_1(t)$ in (\ref{69}) always keeps as zero. This shows that $E'$ is one subset of the largest invariant set of the system with the switching control (\ref{69}). On the other hand, for the switching Lyapunov control (\ref{69}), the system state will keep transferring toward the target state when the state  $\rho(\tilde{0})$ at the zero point $\tilde{0}$ of the bang-bang Lyapunov control does not satisfy the switching condition (\ref{40}); while when (\ref{40}) is satisfied, the control field will switch to the standard Lyapunov control and the system state will still continue evolving toward the target state. This shows that the switching control (\ref{69}) will not generate any new stable state except $\rho_f$ and $\rho_2$ such that the system stops evolving. Hence, the largest invariant set of the system with the switching control (\ref{69}) is still $E'=\{\rho_f\}\cup\{\rho_2\}$.\newline\newline
Conclusion $\mathrm{ii})$. If $\rho_0=\rho_f$, then the conclusion naturally holds. Next, we consider the case of $\rho_0\ne\rho_f$. The condition $\mathrm{tr}(\rho_0\rho_f)\ne0$ implies that $\rho_0\ne\rho_2$. The Lyapunov function (\ref{13}) takes the maximal value $p_2$ when $\rho=\rho_2$ and monotonically decreases under the action of the switching Lyapunov control (\ref{69}). Using Conclusion $\mathrm{i})$, we know that the system trajectory starting from $\rho_0$ necessarily converges to $\rho_f$ contained in the largest invariant set $E'$.\newline\newline
%when $\rho_0\ne\rho_f$ and $\mathrm{tr}(\rho_0\rho_f)\ne0$ hold.
Conclusion $\mathrm{iii})$. The condition $\mathrm{tr}(\rho_0\rho_f)=0$ implies that $\rho_0=\rho_2$. In this case, the use of the initial control $u_1 (t)=S \sin⁡(\omega_{21} t)\,(t\in[0,t'])$ leads to $\mathrm{tr}(\rho(t')\rho_f)\ne0$. Then, Conclusion $\mathrm{ii})$ guarantees that the system trajectory starting from $\rho(t')$ necessarily converges to $\rho_f$.\quad$\blacksquare$
\end{pf}
%For the two-level system (\ref{2level}), the switching control law (\ref{69}) guarantees that the high-frequency oscillation condition in Theorem \ref{thm5} no longer holds. Thus, the bang-bang Lyapunov control in the switching control (\ref{69}) can always drive the system state towards the target state. At the same time, Theorem \ref{thm3} or Theorem \ref{thm4} guarantees that the standard Lyapunov control as a continuously differentiable function can always achieve convergence to the target state. Therefore, the switching control law achieves convergence of the whole system.

%For the switching control between bang-bang Lyapunov controls with different strengths, before the first high-frequency oscillation phenomenon occurs, the switching control (\ref{71}) automatically uses a new bang-bang control strength according to (\ref{70a}) or (\ref{70b}), which means that the high-frequency oscillation condition in Theorem \ref{thm5} does not hold. Thus, Theorem \ref{thm5} guarantees that the system continues evolving towards the target state with the new control strength. Before the second high-frequency oscillation phenomenon occurs, the switching control (\ref{71}) automatically uses the second new control strength according to (\ref{70a}) or (\ref{70b}) again. The system further transfers towards the target state. The above process is repeated, and the system will converge to the target state.

\section{Approximate bang-bang Lyapunov control}
In Section 4, we proposed two switching design methods between Lyapunov control schemes to achieve improved performance. However, the switching points are not easy to determine for a general case. In this section, we further propose two approximate bang-bang (ABB) Lyapunov control approaches that can achieve rapidly convergent control \cite{c35}.

\subsection{ABB Lyapunov control design}
The first ABB Lyapunov control law is designed as
\begin{equation}\label{19}
u_k(T_k)=\frac{2S_k}{1+e^{\gamma_kT_k}}-S_k,\;(k=1,\cdots,m),
\end{equation}
where $S_k>0$ is the maximum admissible strength of the control field $u_k$, and $\gamma_k>0$ is a parameter used to adjust the hardness of the control function. The bigger $\gamma_k$ is, the harder the characteristic of $u_k(T_k)$ is. As $\gamma_k\rightarrow+\infty$, the characteristic of $u_k(T_k)$ approaches the bang-bang Lyapunov control in (\ref{18}).

The second ABB Lyapunov control law is designed as
\begin{equation}\label{20}
u_k(T_k)=\frac{-S_kT_k}{|T_k|+\eta_k},\;(k=1,\cdots,m),
\end{equation}
where $S_k>0$ is the maximum admissible strength of the control field $u_k$, and $\eta_k>0$ is a parameter used to adjust the hardness of the control function. Here, the smaller $\eta_k$ is, the harder the characteristic of $u_k(T_k)$ is. Likewise, as $\eta_k\rightarrow0^+$, the characteristic of $u_k(T_k)$ approaches the bang-bang Lyapunov control in (\ref{18}). The two smooth control laws (\ref{19}) and (\ref{20}) can show similar characteristics to bang-bang Lyapunov control by choosing appropriate hardness parameters. Hence, we call them approximate bang-bang (ABB) Lyapunov controls.

The ABB Lyapunov control laws (\ref{19}) and (\ref{20}) are two special forms of the smooth control law (\ref{17a}). Therefore, the convergence results in Theorems \ref{thm2}, \ref{thm3}, and \ref{thm4} naturally hold for these ABB control laws. That is to say, with the conditions in Theorem \ref{thm3} or Theorem \ref{thm4}, the corresponding ABB Lyapunov control laws (\ref{19}) and (\ref{20}) are always stable.

\subsection{Further construction of $P$}
Since the control functions (\ref{19}) and (\ref{20}) are continuously differentiable, the qualitative construction relation of $P$ in Theorems \ref{thm3} and \ref{thm4} can guarantee convergence to the target state. In order to speed up the control process in the early stages, we can design diagonal elements of $P$ such that the early-stage control has a bang-bang property.

Without loss of generality, we assume that the target state is the $N$-th eigenstate, i.e., $\rho_f=\rho_N$. Consider the construction relation of $P$ in Theorems \ref{thm3} and \ref{thm4}, we denote $P, \rho, H_0$ and $H_k$ as the following block forms:
\begin{gather}
P\triangleq\begin{bmatrix}
pI_{N-1}&0\\
0&p_f
\end{bmatrix}=\begin{bmatrix}
pI_{N-1}&0\\
0&p_N
\end{bmatrix},\label{34a}
\end{gather}
\begin{gather}
\rho\triangleq\begin{bmatrix}
\rho^{\#1}&\rho^{\#2}\\
(\rho^{\#2})^\dag&\rho_{ff}
\end{bmatrix}=\begin{bmatrix}
\rho^{\#1}&\rho^{\#2}\\
(\rho^{\#2})^\dag&\rho_{NN}
\end{bmatrix},\label{34b}\\
H_0\triangleq\begin{bmatrix}
H_0^{\#1}&0\\
0&\lambda_N
\end{bmatrix},\label{34c}\\
H_k\triangleq\begin{bmatrix}
H_k^{\#1}&H_k^{\#2}\\
(H_k^{\#1})^\dag&(H_k)_{NN}
\end{bmatrix}\triangleq\begin{bmatrix}
H_k^{\#1}&R_k\\
R_k^\dag&(H_k)_{NN}
\end{bmatrix},\label{47}
\end{gather}
where $I_{N-1}$ is identity matrix of order $N-1$; $\rho^{\#1}, H_0^{\#1}$, and $H_k^{\#1}$ are Hermitian matrices of order $N-1$, and $H_0^{\#1}=\text{diag}[\lambda_1,\cdots,\lambda_{N-1}]$; $\rho^{\#2}$ and $R_k=H_k^{\#2}$ are column vectors of dimension $N-1$; and $\rho_{NN}=\rho_{ff}$ and $(H_k)_{NN}$ are real numbers.

Calculating $T_1(t)$ gives
\begin{eqnarray}\label{35a}
T_k(t)&=&-i(p-p_N)\mathrm{tr}\!\!\left(
\begin{bmatrix}
\rho^{\#1}(t)&\rho^{\#2}(t)\\
(\rho^{\#2}(t))^\dag&\rho_{NN}(t)
\end{bmatrix}
\begin{bmatrix}
0&R_k\\
-R_k^\dag&0
\end{bmatrix}
\right)\nonumber\\
&=&-2(p-p_N)\cdot\Im\,\langle R_k|\rho^{\#2}(t)\rangle.
\end{eqnarray}
Since the evolution state starting from any initial pure state always stays, we can obtain an estimate of the amplitude of $T_k(t)$ as:
\begin{align}\label{48}
\quad|T_k(t)|&=2(p-p_N)\cdot|\Im\,\langle R_k|\rho^{\#2}(t)\rangle|\nonumber\\
&\leq2(p-p_N)\cdot|\langle R_k|\rho^{\#2}(t)\rangle|\nonumber\\
&\leq2(p-p_N)\cdot\|R_k\|\cdot\|\rho^{\#2}(t)\|\nonumber\\
&=2(p-p_N)\cdot\|R_k\|\cdot\nonumber\\
&\quad\quad\sqrt{\rho_{11}\rho_{NN}+\cdots+\rho_{N-1,N-1}\rho_{NN}}\nonumber\\
&=2(p-p_N)\cdot\|R_k\|\cdot\sqrt{(1-\rho_{NN}(t))\rho_{NN}(t)}\nonumber\\
&\leq(p-p_N)\cdot\|R_k\|.
\end{align}
For (\ref{48}), when $\text{tr}(\rho_0\rho_f)<\frac{1}{2}$, $|T_k(t)|$ can reach the maximum value $(p-p_N)\|R_k\|$ during the evolution process.

Assume that the initial state $\rho_0$ satisfies $\text{tr}(\rho_0\rho_f)<\frac{1}{2}$ and that the control laws (\ref{19}) and (\ref{20}) are regarded as having the bang-bang property when the control value reaches $\beta S,(\beta\approx1,\beta<1)$. In this case, we calculate from (\ref{19}) and (\ref{20}) that $|T_k|=\frac{1}{\gamma_k}\ln\frac{1+\beta}{1-\beta}$ and $|T_k|=\frac{\beta}{1-\beta}\eta_k$. Thus, when $p-p_f$ satisfies $(p-p_f )\|R_k\|>\frac{1}{\gamma_k}\ln\frac{1+\beta}{1+\beta}$ and $(p-p_f)\|R_k\|>\frac{\beta}{1-\beta}\eta_k$, respectively, the bang-bang property will be dominant in the control process. These two expressions imply that, when $\gamma_k$ is relatively small or $\eta_k$ is relatively large, $p-p_f$ should be large, and vice versa. The selection can also be explained as follows. When $\gamma_k$ is very large or $\eta_k$ is very small, a large $p-p_f$ will put $|T_k|$ in the saturation regions of the functions (\ref{19}) and (\ref{20}). This makes the whole control process similar to the bang-bang Lyapunov control in (\ref{18}).

\section{Numerical Examples}
In this section, we present three numerical examples to demonstrate the performance of the proposed rapid Lyapunov control strategies. In the first example of two-level quantum systems, we compare the standard Lyapunov control, the switching Lyapunov control and ABB Lyapunov control. In the other two examples for three-level quantum systems and two-qubit superconducting systems, the ABB Lyapunov control strategies are compared with the standard Lyapunov control. These simple quantum systems in the examples have significant applications in quantum information technology. The ABB Lyapunov control strategies are also applicable to general $N$-level quantum systems although more control fields are usually required to guarantee the controllability.
\subsection{Two-level quantum system}
Consider the two-level system (\ref{2level}) where $H_0=\left[\begin{smallmatrix}
0.4&0\\
0&0
\end{smallmatrix}\right]$ and $H_1=\left[\begin{smallmatrix}
0&1\\
1&0
\end{smallmatrix}\right]$; the maximum admissible strength of the control field is $S=0.2$; the initial pure state and the target eigenstate are given as $\rho_0=\frac{1}{6}\left[\begin{smallmatrix}
1&\sqrt{5}\\
\sqrt{5}&5
\end{smallmatrix}\right]$ and $\rho_f=\left[\begin{smallmatrix}
1&0\\
0&0
\end{smallmatrix}\right]$, respectively. According to Theorem \ref{thm3} or Theorem \ref{thm4}, we set $P=\text{diag}[0.5,1]$.

\begin{figure}
\begin{center}
\includegraphics[width=8.5cm]{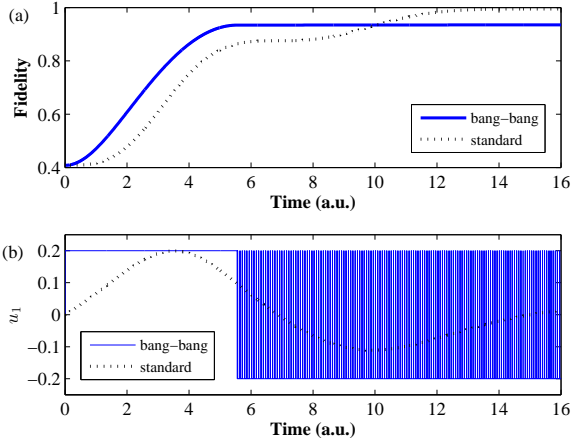}    % The printed column
\caption{The evolution curves of the fidelities with the target state (Fig. 1 (a)) and the control fields (Fig. 1 (b)) under the bang-bang Lyapunov control and the standard Lyapunov control, where the bang-bang Lyapunov control shows a high-frequency oscillation phenomenon.}  % width is 8.4 cm.
\label{Fig1}                                 % Size the figures
\end{center}                                 % accordingly.
\end{figure}

\begin{figure}
\begin{center}
\includegraphics[width=8.5cm]{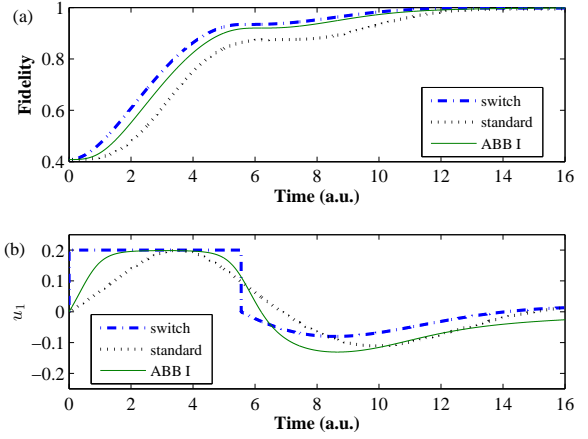}    % The printed column
\caption{The evolution curves of fidelities with the target state (Fig. 2(a)) and the control fields (Fig. 2(b)) under the switching Lyapunov control (\ref{69}) and the ABB Lyapunov control (\ref{19}) (ABB I).}  % width is 8.4 cm.
\label{Fig22}                                 % Size the figures
\end{center}                                 % accordingly.
\end{figure}

We present simulation results first for the bang-bang Lyapunov control (\ref{18}) and standard Lyapunov control (\ref{17b}); then for the switching control (\ref{69}). In order to compare the control effect of the two classes of rapid Lyapunov control methods, we also present the simulation results for the approximate bang-bang Lyapunov control (\ref{19}) with $k=1$ in Section 5. In simulations, we let the control gain in (\ref{17b}) and (\ref{69}) be $K_1=0.4$ such that the maximum value of the standard Lyapunov control can reach up to the maximum admissible strength $S=0.2$. Through multiple simulations, we choose $\gamma_1=11$ in (\ref{19}) such that the ABB Lyapunov control (\ref{19}) can achieve rapid convergence. Simulation results are shown in Fig. \ref{Fig1} and Fig. \ref{Fig22}.

Fig. \ref{Fig1} shows the evolution curves of fidelities with the target state and the control fields under the bang-bang Lyapunov control and the standard Lyapunov control. It can be seen from Fig. \ref{Fig1}(a) that the standard Lyapunov control achieves convergence to the target state; and that the bang-bang Lyapunov control has better rapidness in the early stages but a high-frequency oscillation phenomenon occurs from $t=5.5$ as shown in Fig. \ref{Fig1}(b).

Fig. \ref{Fig22} shows the evolution curves of fidelities with the target state, and the control fields under the switching Lyapunov control, the approximate bang-bang Lyapunov control, and the standard Lyapunov control. It can be seen from Fig. \ref{Fig22} that these two classes of rapid Lyapunov controls have similar control performance and achieve excellent convergence to the target state. They have better rapidness than the standard Lyapunov control. In addition, the switching Lyapunov control has slightly better rapidly convergence than the ABB Lyapunov control in (\ref{19}). This is because that the switching Lyapunov control (\ref{69}) always keeps the maximal control value 0.2 in the early stages (see Fig. \ref{Fig22}(b)). More simulation results also show that it is possible in the switching Lyapunov control to obtain better performance by using a control strength less than $S$ for the initial bang-bang control field.

\subsection{$\Xi$-type three-level system}
Consider a three-level system with $\Xi$-type configuration. This system is controlled by only one control field, and its internal and control Hamiltonians are given as
$H_0=\left[\begin{smallmatrix}
0&0&0\\
0&0.3&0\\
0&0&0.9
\end{smallmatrix}\right]$ and $H_1=\left[\begin{smallmatrix}
0&1&0\\
1&0&1\\
0&1&0
\end{smallmatrix}\right]$, respectively. Assume that the maximum admissible strength of the control field is 0.1, and that the target state is the second eigenstate of the system, i.e., $\rho_f=|\lambda_2\rangle\langle\lambda_2|$. Based on Theorems \ref{thm3} and \ref{thm4}, $\rho_f$ is isolated in the invariant set, and any system trajectory starting from initial pure states necessarily converges to $\rho_f$ under the action of the control law (\ref{17a}) or (\ref{33}) with the control function forms of (\ref{19}) or (\ref{20}).

\begin{figure}
\begin{center}
\includegraphics[width=8.5cm]{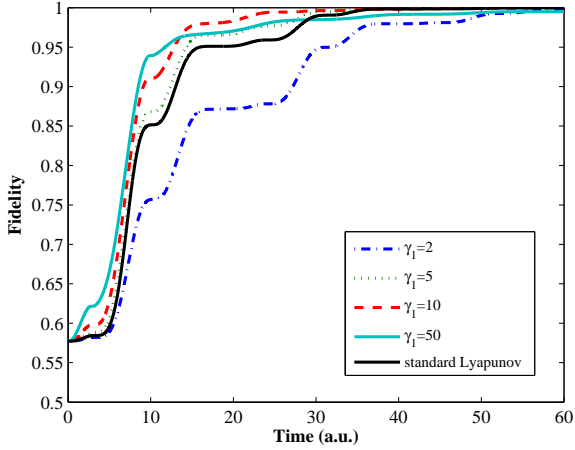}    % The printed column
\caption{The time evolution of the fidelities under the standard Lyapunov control (\ref{17b}) and the approximate bang-bang Lyapunov control (\ref{19})
with different hardness parameter values.}  % width is 8.4 cm.
\label{Fig3}                                 % Size the figures
\end{center}                                 % accordingly.
\end{figure}

Assume that that the initial state is given as
$\rho_0=\frac{1}{3}\left[\begin{smallmatrix}
1&1&1\\
1&1&1\\
1&1&1
\end{smallmatrix}\right]$. In order to compare with the standard Lyapunov control (\ref{17b}), we choose $K_1=0.155$ in (\ref{17b}) so that the maximum strength of the standard Lyapunov control can reach the maximum admissible strength of the control field 0.1. Based on Theorem \ref{thm3}, we set $P=\text{diag}\,[1,0.5,1]$, choose (\ref{19}) as the control law and set its hardness parameter as $\gamma_1=2,5,10,50$, respectively. The simulation results are shown in Fig. \ref{Fig3}.

It can be seen from Fig. \ref{Fig3} that, as the hardness parameter increases, the rapidness of the approximate bang-bang Lyapunov control (\ref{19}) is reinforced. At the same time its convergence rate decreases in the later stages. Based on simulation experiments, when $5\leq\gamma_1\leq10$, we can obtain improved control performance compared with the standard Lyapunov control for this system, considering both rapidness and convergence.

\subsection{Two-qubit superconducting system}
In this subsection, we consider the control problem of a superconducting quantum system of two qubits. Superducting qubits have been recognized as promising quantum information processing units due to their scalability and design flexibility \cite{c36,c37,c38}. Superconducting qubits can behave quantum mechanically while they can be controlled by adjusting some classical quantities such as currents and voltages. Here we consider the coupled two-qubit system in \cite{c39} where two charge qubits are coupled via an LC-oscillator. The system model can be described as
\begin{align}\label{72}
\dot{\rho}(t)&=-i\big[u_{1z}\sigma_z^{(1)}\otimes I_2+u_{2z}I_2\otimes\sigma_z^{(2)}+u_{1x}\sigma_x^{(1)}\otimes I_2\nonumber\\
&\quad\quad+u_{2x}I_2\otimes\sigma_x^{(2)}+u_{xx}\sigma_x^{(1)}\otimes\sigma_x^{(2)},\;\rho(t)\big],
\end{align}
where $I_2=\left[\begin{smallmatrix}
1&0\\
0&1
\end{smallmatrix}\right]$; $\sigma_z^{(1)}=\sigma_z^{(2)}=\left[\begin{smallmatrix}
1&0\\
0&-1
\end{smallmatrix}\right]$ and $\sigma_x^{(1)}=\sigma_x^{(2)}=\left[\begin{smallmatrix}
0&1\\
1&0
\end{smallmatrix}\right]$ are the Pauli matrices along the $z$ and $x$ directions, respectively. Considering the experimental parameters \cite{c40}, we assume that these control fields satisfy the following constraints: $1\,\mathrm{GHz}\leq|u_{jz}(t)|\leq20\,\mathrm{GHz}$, $|u_{jx}(t)|\leq10\,\mathrm{GHz},(j=1,2)$; and $|u_{xx}(t)|\leq0.5\,\mathrm{GHz}$.

\begin{figure}
\begin{center}
\includegraphics[width=8.5cm]{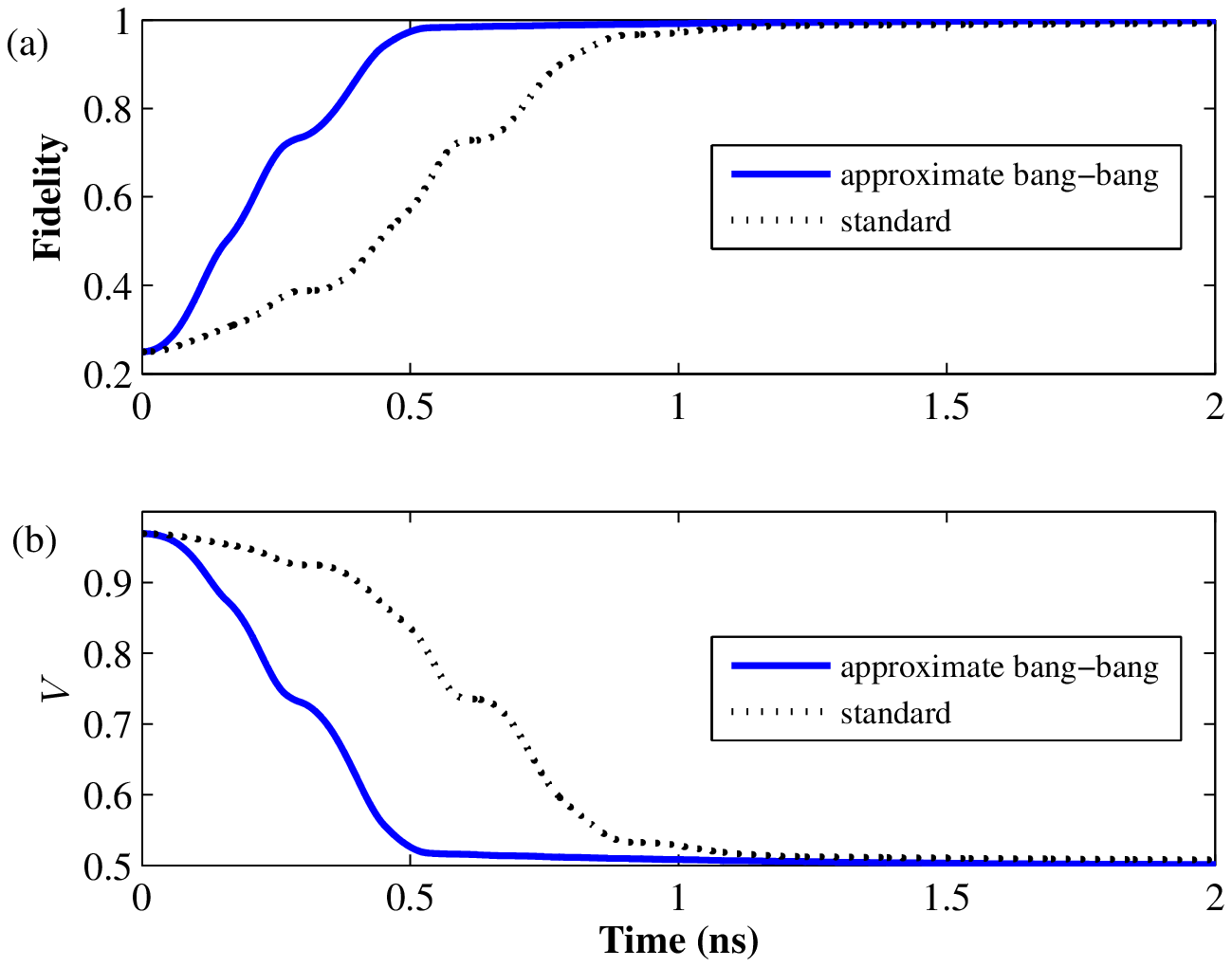}    % The printed column
\caption{The evolution curves of the fidelities (Fig. 4 (a)) and the Lyapunov functions (Fig. 4 (b)) under the approximate bang-bang Lyapunov control (\ref{20}) and the standard Lyapunov control (\ref{17b}) on a two-qubit superconducting system.}  % width is 8.4 cm.
\label{Fig4}                                 % Size the figures
\end{center}                                 % accordingly.
\begin{center}
\includegraphics[width=8.5cm]{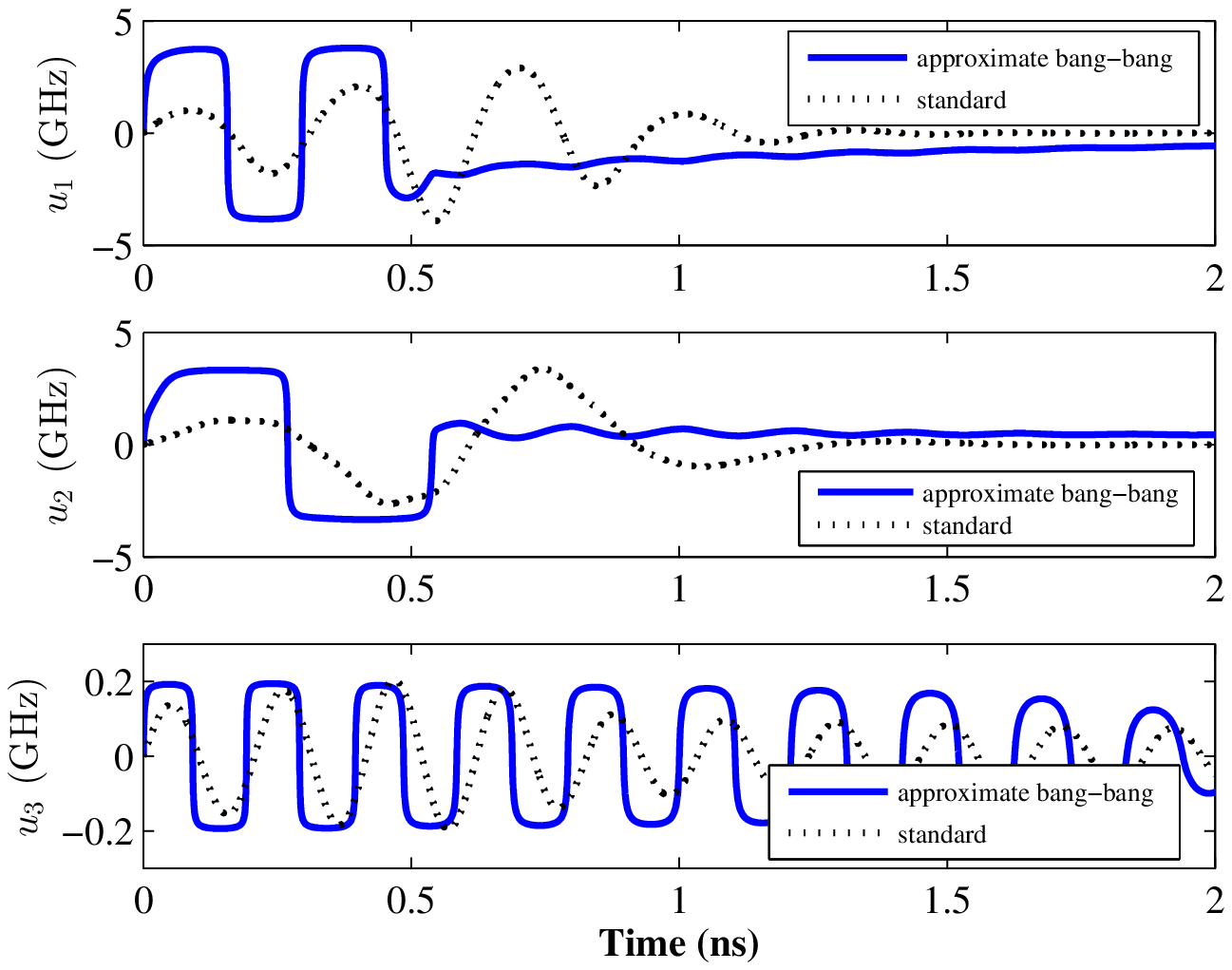}    % The printed column
\caption{The evolution curves of the three control fields under the approximate bang-bang (ABB) Lyapunov control (\ref{20}) and the standard Lyapunov control (\ref{17b}) on a two-qubit superconducting system, in which the blue solid lines are the evolution curves of the ABB Lyapunov control, and the black dotted lines correspond to the standard Lyapunov control.}  % width is 8.4 cm.
\label{Fig5}                                 % Size the figures
\end{center}                                 % accordingly.
\end{figure}

Assume that the target state is $\rho_f=|\lambda_1\rangle\langle\lambda_1|$. For simplicity, we let $u_{1z}(t)=10\,\mathrm{GHz}$, $u_{2z}(t)=5\,\mathrm{GHz}$, $u_{1x}(t)=u_1(t)$, $u_{2x}(t)=u_2(t)$, and $u_{xx}(t)=u_3(t)$. Thus, the model (\ref{72}) can be written as
\begin{align}\label{73}
\dot{\rho}(t)=-i\Big[H_0+\sum_{k=1}^3u_k(t)H_k,\;\rho(t)\Big],
\end{align}
where $H_0=\text{diag}\,[15,5,-5,-15]$, $H_1=\sigma_x^{(1)}\otimes I_2$, $H_2=I_2\otimes\sigma_x^{(2)}$, $H_3=\sigma_x^{(1)}\otimes\sigma_x^{(2)}$, $|u_j(t)|\leq10\,\mathrm{GHz}$, $(j=1,2)$, and $|u_3(t)|\leq0.5\,\mathrm{GHz}$. The initial state is given as the pure state $\rho_0=\frac{1}{16}\left[\begin{smallmatrix}
1&1&1&\sqrt{13}\\
1&1&1&\sqrt{13}\\
1&1&1&\sqrt{13}\\
\sqrt{13}&\sqrt{13}&\sqrt{13}&13
\end{smallmatrix}\right]$.

Set $P=\text{diag}\,[0.5,1,1,1]$, and choose (\ref{20}) as our control law. For comparison with the standard Lyapunov control, we choose $K_1=15$, $K_2=12$, and $K_3=0.6$ for the standard Lyapunov control (\ref{17b}). According to simulation results, we know that the maximum values of the control fields are 3.9, 3.4, and 0.2, respectively. Therefore, we set these three values to be the maximum admissible strengths of the three approximate bang-bang Lyapunov control fields, i.e., $S_1=3.9$, $S_2=3.4$, and $S_3=0.2$. Further, we choose their hardness parameters as $\eta_1=\eta_2=0.005$, and $\eta_3=0.01$, respectively. The simulation results are shown in Fig. \ref{Fig4} and Fig. \ref{Fig5}.

It can be seen from Fig. \ref{Fig4} that both the approximate bang-bang Lyapunov control (\ref{20}) and the standard Lyapunov control (\ref{17b}) achieve convergence, but the rapidness of the approximate bang-bang Lyapunov control (\ref{20}) is better than that of the standard Lyapunov control (\ref{17b}). Fig. \ref{Fig5} shows that the three control fields associated with the approximate bang-bang Lyapunov control (\ref{20}) have a bang-bang like property in the early stages of the control, which speeds up the control process.

\section{Conclusions}
In this paper, we have designed a switching Lyapunov control strategy between the bang-bang Lyapunov control and standard Lyapunov control and two approximate bang-bang Lyapunov control laws. These Lyapunov control laws can achieve rapidly convergent control for quantum systems by choosing appropriate parameters. In particular, convergence has been analyzed via the LaSalle invariance principle, and a construction method for the degrees of freedom in the Lyapunov function has been provided. We have also derived a sufficient condition for the existence of high-frequency oscillations that can be used for switching Lyapunov control design. Simulation experiments showed that the proposed Lyapunov control methods can achieve improved performance for manipulating quantum systems. Further research includes optimizing these parameters in the control laws and extending the proposed methods to more general quantum systems.

\section*{Appendix: Switching between bang-bang Lyapunov controls with different strengths}
Here, we can design another switching Lyapunov control strategy for two-level quantum systems in Section 4, i.e., switching between bang-bang Lyapunov controls with different control bounds.

Observing the high-frequency oscillation condition for two-level systems in Theorem \ref{thm5}, it is clear that reducing the bang-bang Lyapunov control strength can avoid high-frequency oscillations. This observation inspires us to develop a new switching design strategy involving switching between bang-bang Lyapunov controls with different control bounds.

Assume that the state $\rho(\tilde{0})$ at the zero point $\tilde{0}$ of $T_1(t)$ satisfies the high-frequency oscillation condition in the bang-bang Lyapunov control (\ref{18}). In theory, any positive number such that the infinitesimal high-frequency oscillation condition is not satisfied may be chosen as the strength of the new bang-bang Lyapunov control. The selection of strengths is not unique. For instance, we can design the following bang-bang control strength:
\begin{equation}\label{70a}
S(\tilde{0}^+)=\frac{\mu_1\omega_{12}|\rho_{12}(\tilde{0})|}{|r|(\rho_{11}(\tilde{0})-\rho_{22}(\tilde{0}))},
\end{equation}
where $S(\tilde{0}^+)$ represents the strength of the new bang-bang Lyapunov control after time $\tilde{0}$; and $\mu_1\in(0,1)$ is a constant guaranteeing that the new control strength does not satisfy the high-frequency oscillation condition.

Considering the fact that $0\leq2\sqrt{(1-\rho_{11}(t))\rho_{11}(t)}=2|\rho_{12}(t)|\leq1$ and $|\rho_{12}(t)|\rightarrow 0$ as $t\rightarrow\infty$, we may design a coefficient-varying bang-bang control strength as follows:
\begin{equation}\label{70b}
S(\tilde{0}^+)=\frac{2\mu_2\omega_{12}|\rho_{12}(\tilde{0})|^2}{|r|(\rho_{11}(\tilde{0})-\rho_{22}(\tilde{0}))},
\end{equation}
where the constant $\mu_2\in(0,1)$ can be properly chosen. It is clear that the coefficient $2\mu_2|\rho_{12}(\tilde{0})|$ also guarantees that the high-frequency oscillation condition does not hold. Hence, we can use the following switching control law for the two-level system (\ref{2level}):
\begin{equation} \label{71}
u_1(t)=\left\{ \begin{aligned}
         &-S(\tilde{0}^+)\cdot\mathrm{sgn}(T_1(t)), \Big(\frac{|r|(\rho_{11}(\tilde{0})-\rho_{22}(\tilde{0}))}{|\rho_{12}(\tilde{0})|}\\
         &\hspace{9em}\geq\frac{\omega_{12}}{S(\tilde{0}^-)}, \rho_{12}(\tilde{0})\ne0\Big)\\
         &-S(\tilde{0}^-)\cdot\mathrm{sgn}(T_1(t)),\;\mathrm{otherwise}
         \end{aligned} \right.
\end{equation}
where $S(\tilde{0}^-)$ represents the last strength of the bang-bang Lyapunov controls before time $\tilde{0}$. In particular, the strength of the bang-bang Lyapunov control before the first zero point is $S$.

\begin{ack}                               % Place acknowledgements
This work was supported by the Fundamental Research Funds for the Central Universities (No. WK2100100019, China), the National Natural Science Foundation of China (No. 60904033), and the Australian Research Council (DP130101658, FL110100020). % here.
\end{ack}

%\bibliographystyle{plain}        % Include this if you use bibtex
%\bibliography{autosam}           % and a bib file to produce the
                                 % bibliography (preferred). The
                                 % correct style is generated by
                                 % Elsevier at the time of printing.

%\appendix{Appendix: A lemma}
%\begin{lem}\label{prop3}
%Let $F=F(x)$ be a continuous function in the variable $x$, and satisfy: 1) $F(0)=0$; and 2) the derivative at point 0 is not equal to zero, i.e., $F'(0)\ne0$. Then, there exists a small %positive number $\alpha_1>0$ such that $F(x)$ and $F'(0)$ have the same sign in the interval $(0,\alpha_1)$.
%\end{lem}

\end{document}